\documentclass[aps,prd,floatfix,superscriptaddress,showpacs,showkeys]{revtex4}

\usepackage{amssymb,amsmath,amsfonts,latexsym,graphicx,epsfig}

\RequirePackage{xspace}

\usepackage{titlesec}
\usepackage{color}
\titleformat{\section}{\large\bfseries}{\thesection}{1em}{}

\newcommand{\Jpsi}{\ensuremath{J\!/\!\psi}\xspace}

\newcommand{\bea}{\begin{eqnarray}}
\newcommand{\ena}{\end{eqnarray}}

\newcommand{\be}{\begin{equation}}
\newcommand{\en}{\end{equation}}

\newcommand{\nn}{\nonumber\\}

\newcommand{\bl}{\bigl}
\newcommand{\br}{\bigr}
\newcommand{\la}{\langle}
\newcommand{\ra}{\rangle}

\newcommand{\Tr}{\mbox{\rm{tr}}}

\begin{document}
        
\title{On the meson mass spectrum
in the covariant confined quark model}

\author{Gurjav Ganbold}
\affiliation{Bogoliubov Laboratory of Theoretical Physics, \\
Joint Institute for Nuclear Research, 141980 Dubna, Russia}
\affiliation{Institute of Physics and Technology, 210651, Ulaanbaatar,
Mongolia}
\author{Thomas Gutsche}
\affiliation{Institut f\"ur Theoretische Physik, Universit\"at T\"ubingen,\\
Kepler Center for Astro and Particle Physics,\\
Auf der Morgenstelle 14, D-72076, T\"ubingen, Germany}

\author{Mikhail A. Ivanov}
\affiliation{Bogoliubov Laboratory of Theoretical Physics, \\
Joint Institute for Nuclear Research, 141980 Dubna, Russia}

\author{Valery E. Lyubovitskij}
\affiliation{Institut f\"ur Theoretische Physik, Universit\"at T\"ubingen,\\
Kepler Center for Astro and Particle Physics,\\
Auf der Morgenstelle 14, D-72076, T\"ubingen, Germany}
\affiliation{Department of Physics, Tomsk State University,
634050 Tomsk, Russia}
\affiliation{Mathematical Physics Department,
Tomsk Polytechnic University, \\ 
Lenin Avenue 30, 634050 Tomsk, Russia}

\begin{abstract}

We provide a new insight into the problem of generating the
hadron mass spectrum in the framework of the covariant confined quark model.
One of the underlying principles of this model is the compositeness
condition which means that the wave function renormalization constant
of the elementary hadron is equal to zero. In particular, this equation
allows to express the Yukawa coupling of the meson fields to
the constituent quarks as a function of other model parameters.
In addition to the compositeness condition  we also employ
a further equation which relates the meson mass function to
the Fermi coupling.
Both equations guarantee that the Yukawa-type theory is equivalent to
the Fermi-type theory thereby providing an interpretation of the meson
field as the bound state of its constituent fermions (quarks).
We evaluate the Fermi-coupling as a function of meson
(pseudoscalar and vector) masses and vary the values of the masses
in such a way to obtain a smooth behavior for the resulting curve.
The mass spectrum obtained in this manner is found to be in good agreement
with the experimental data.
We also compare the behavior of our Fermi-coupling with the strong QCD
coupling $\alpha_s$ calculated in an QCD-inspired approach.

\end{abstract}

\pacs{12.39.Ki,13.30.Eg,14.20.Jn,14.20.Mr}
\keywords{relativistic quark model, light and heavy mesons,
mass spectrum and decay constants}

\maketitle

%=================================================================

\section{Introduction}

One of the puzzles of hadron physics is the origin of the hadron masses.
The Standard Model (SM) and, in particular, quantum chromodynamics (QCD)
operate only with fundamental particles (quarks, leptons, neutrinos),
gauge bosons and the Higgs. It is not yet clear how to explain the appearance
of the multitude of observed hadrons and elucidate the generation
of their masses.
Therefore, the calculation of the hadron mass spectrum in a quality
comparable to the precision of experimental data still remains one of the
major problems in QCD.

Actually, even before QCD was set up as the fundamental theory of
strong interactions, it was understood that
it is a difficult problem to describe a composite particle
within quantum field theory as based on the relativistic S-matrix.
The reason is that quantum field theory operates with free fields
which are quantized by imposing commutator (anti-commutator)
relations between creation and annihilation operators.
The asymptotic in- and out-states are constructed by means of
these operators acting on the vacuum state. Physical processes are 
described by the elements of the S-matrix taken for the relevant in- and
out-states. In perturbation theory, which originally has
a mathematical meaning, the matrix elements are represented by a set 
of the Feynman diagrams
which are the convolution of free Green functions (or propagators).
The original Lagrangian describing free fields and their interactions
requires renormalization, i.e. the transition from bare or
unrenormalized quantities like mass, wave function, coupling constant
to the physical or renormalized ones. In particular, the bare field
is related to the dressed one by the wave function
renormalization constant as $\phi_0=Z^{1/2}\phi_r$. One can see that
the bare field $\phi_0$ may be eliminated from the Lagrangian
by putting its wave function renormalization constant to be zero,
$Z=0$.  Probably one of the first who suggested to use the equation
$Z=0$ as a compositeness condition was Jouvet \cite{Jouvet:1956ii}.
He showed that the four-fermion theory is equivalent to a Yukawa-type theory
if the renormalization constant of the boson field is set to zero.
The crucial point in comparison of the two theories is related
to the renormalization of the Yukawa-type theory, i.e.
the transition from the bare quantities
(boson mass, boson wave function, Yukawa coupling) to the physical or
renormalized ones. Then the renormalized (experimental) boson mass
and Yukawa coupling may be expressed through the Fermi constant via
the compositeness condition.

Salam extended the condition of setting the wave function renormalization
constant to zero by requiring that the vertex renormalization constant
must be equal to zero, too \cite{Salam:1962ap}. In this way he showed
that the obtained theory included the bootstrap idea in quantum field theory.
Some aspects of this approach were developed further in \cite{Braun:1970df}.

Weinberg showed that the compositeness condition $Z=0$ results in
a sum rule for the coupling of a composite particle to its constituents as
a function of energy \cite{Weinberg:1962hj}. 
Further developments and 
applications of  the compositeness condition may be found in 
a monograph~\cite{Lurie:1968zz} and a review \cite{Hayashi:1967hk}. 
We also should mention the work of \cite{Anikin:1995cf}. 
There it was shown that
the nonlocal Nambu--Jona-Lasinio (NJL) model is equivalent
to the Yukawa theory of scalar and pseudoscalar fields interacting with
their constituents if the wave function normalization constants
are set to zero. Some physical, low-energy observables have been
calculated in the framework of this approach.

One of the phenomenological approaches is  the model of induced quark currents.
It is based on the hypothesis that the QCD vacuum is realized by 
the ~anti-!self-dual homogeneous gluon field~\cite{Burdanov:1996uw}.
The confining properties of the vacuum field,
chiral symmetry breaking, and the localization of a composite field
at the center of mass of the quark bound state
can explain the distinctive features of the meson spectrum: mass splitting
between pseudoscalar and vector mesons, Regge trajectories
and the asymptotic mass formulas in the heavy quark limit.
This model describes to within ten percent accuracy the masses and
weak decay constants of mesons.

In a series of 
papers~\cite{Efimov:2001ng,Ganbold:2009ak,Ganbold:2010bu,Ganbold:2012zz}
relativistic models with specific forms of  analytically confined
propagators have been developed to study some aspects of
low-energy hadron physics.
The role of analytic confinement in the formation of two-particle bound states
has been analyzed within a simple Yukawa model of two interacting scalar
fields,  the prototypes of 'quarks' and 'gluons' ~\cite{Efimov:2001ng}.
The spectra of the' two-quark' and 'two-gluon'  bound states have been defined
by using master constraints similar to the ladder Bethe-Salpeter equations.
The 'scalar confinement' model could explain semiquantitatively the
asymptotically linear Regge trajectories of 'mesonic'  excitations and
the existence of massive 'glueball' states. It  has also been shown that
physically reasonable bound  states could be formed at relatively weak 
coupling. An extension of this model by introducing color and spin degrees of
freedom, different constituent quark masses and the confinement size parameter
has been performed in~\cite{Ganbold:2009ak}. Specific forms of analytically
confined propagators of quarks and gluons have been used to take into account
the correct symmetry structure of the quark-gluon interaction in the
confinement region. The masses of conventional mesons have been estimated
(with relative errors less than $ 3.5$ per cent) in a wide energy range.
As a further test the calculated weak decay constants of light mesons
($f_{\pi}$ and $f_K$) are in good accordance with the
experimental data. Additionally,  the lowest-state glueball mass has been
predicted which is in reasonable agreement with other theoretical approaches.
A phenomenological model with infrared-confined propagators has been developed
to take into account the dependence of the QCD effective coupling
$\alpha_s=g^2/4\pi$ on the mass scale and to study its behavior at large
distances ~\cite{Ganbold:2012zz}.  First the pseudoscalar and vector meson
masses  have been calculated in the region above ~2 GeV. By fitting the
estimated masses to the recent experimental data model
parameters could be fixed, namely the constituent quark masses 
$m_f, \{f=u,d,s,c,b\}$ and the value $\Lambda$ of
the confinement scale. By fitting the experimental
masses of intermediate and light mesons  we have predicted the behavior of
$\alpha_s$ in the low-energy domain. Note, we have also derived analytically
a new, specific and finite behavior of $\alpha_s(M)$ at the origin $M=0$
that increased by decreasing the value of the confinement scale value.  
Note that $\alpha_s(0)$ depends on $\Lambda$, we fixed $\alpha_s(0)=0.757$ for
$\Lambda=345$ MeV~\cite{Ganbold:2010bu} and $\alpha_s(0)=0.8498$
for $\Lambda=220$ MeV~\cite{Ganbold:2014zda}.

The compositeness condition $Z_H=0$ is also one of the key ingredients in the
relativistic constituent quark model developed for the first time
in \cite{Efimov:1988yd} (see also the extensive treatment 
in \cite{Efimov:1993ei}).
The model has found numerous applications both in the meson sector
\cite{meson} and in baryon physics \cite{baryon}. In the latter case
baryons are considered as relativistic systems
composed of three quarks. The next step in the development of the model
has been done in  Ref.~\cite{Branz:2009cd} where infrared confinement
was introduced guaranteeing the absence of all possible thresholds
corresponding to quark production. The implementation of
quark confinement allowed to use the same values for
the constituent quark masses both for the simplest quark-antiquark
systems (mesons) and more complicated multiquark configurations (baryons,
tetraquarks, etc.).  Note that we prefer to name this approach as
{\it the covariant quark model} which is more appropriate in the
context of a comparison with other quark models.

The infrared cutoff parameter $\lambda$ is taken
to have a common value for all processes considered. The model parameters
(constituent quark masses $m_q$, the infrared cutoff parameter $\lambda$
and the size parameters $\Lambda_H $ that characterize the distribution
of the constituent quarks inside the hadron $H$) have been determined
by a fit to available experimental data. The last fit was performed in
Ref.~\cite{Ivanov:2011aa}. This approach was successfully applied
in the calculation of transition form factors needed to study
the semileptonic, nonleptonic  and rare decays of the $B$-meson
and the $\Lambda_b$-baryon \cite{B-physics}.
The X(3872) meson was treated as four-quark state (tetraquark)
\cite{tetraquark}. Its strong and radiative decays have been calculated.
For reasonable values of the size parameter of
the X(3872) we found consistency with the available experimental data.

It should be emphasized that the experimental values of hadron masses are used
in all calculations performed in the covariant quark model.
As discussed above,  the theory with Yukawa couplings describing
the interaction of mesons and their constituent quarks is equivalent
to the theory with a four-fermion interaction. This is the case if, first,
the wave function
renormalization constant of the meson field is equal to zero and, second,
the coupling $G$ characterizing the strength of the  four-fermion interaction
is related to the meson mass function. Up to now we have used the first
equation $Z_H=1-g^2_r\tilde\Pi'(m^2_H)=0$ to determine the renormalized
Yukawa coupling $g_r$ as a function of meson mass and model parameters.
In this paper we are aiming to use also the second equation
$G\tilde\Pi(m^2_H)=1$ to investigate the dependence of the coupling $G$
on physical meson masses.

The paper is organized  in the following way. In Sec.II we give a brief
sketch of the approach to the bound state problem in quantum field theory
based on the compositeness condition with $Z_H=0$. 
By using the functional integral
we demonstrate explicitly that the four-fermion theory with the Fermi
coupling $G$ is equivalent to the Yukawa-type theory if, first,
the wave function renormalization constant in the Yukawa theory
is equal to zero and, second,  the Fermi coupling $G$
is inversely proportional to the meson mass function calculated
at the physical meson mass. In Sec.III we give the details of the calculation
for the mass function of pseudoscalar and vector mesons in the
framework of the covariant quark model. In Sec.IV we update the fit
of model parameters and calculate the Fermi coupling $G$ as a function
of the physical mass in a quite large region ranging from the $\pi$ up to
$B_c$ mesons. We suggest {\it a smoothness criterion} to generate
a continuous behavior of the Fermi coupling $G$. The mass spectrum obtained
in this manner is found to be in good agreement with the experimental data.
We compare the behavior of $G$ with the strong QCD coupling $\alpha_s$
calculated in the QCD-inspired approach.
Finally, in Sec. IV we summarize our findings.

\section{The compositeness condition $Z_H=0$ }

The use of the compositeness condition $Z_H=0$ is one of the outstanding
approaches to the bound state problem in quantum field theory.
Historically, this condition first appeared when looking for
the bound state in a four-fermion theory with the Lagrangian
\be
{\cal L}_F = \bar q (i \not\! \partial - m_q) q \,+\,
           \frac{G}{2}\, \bl(\bar q\Gamma q\br)^2\,.
\label{eq:Fermi}
\en
Here, for simplicity, we drop all color and flavor indices. For the general
Dirac matrix we use $\Gamma=I, i\gamma^5$, i.e. we restrict to bound states
with zero spin. We will consider the bound state problem by using
the chain (one-loop) approximation but the result is general
and can be proved to all orders of perturbation theory. 
In the following, for simplicity, we will not consider the renormalization 
of the fermion (``quark'') fields.

The bare Lagrangian with a Yukawa coupling of the boson field $\phi_0$ to the
fermions is written as 
\be
{\cal L}_Y =  \bar q (i \not\! \partial - m_q) q \,+\,
            \frac12 \phi_0 (\Box - m^2_0) \phi_0
            + g_0 \phi_0\,\bl( \bar q\Gamma q\br)\,, 
\quad\text{where}\quad
\Box = - \partial^\mu\partial_\mu\,.
\label{eq:Yukawa}
\en

The more transparent way to demonstrate the renormalization
procedure is the use of the functional integral. The vacuum
generating functional for the Yukawa theory is written as
\be
Z_Y = \int\!\! {\cal D}\phi_0\!\int\!\! {\cal D}\bar q\! \int\!\! {\cal D} q\,
e^{i\int\! dx {\cal L}_Y(x)}\,.
\label{eq:Yukawa_func1}
\en
 Hereafter, we will drop all irrelevant normalization constants.
 Integrating out the quark fields
\bea
&&
\int\!\! {\cal D}\bar q\! \int\!\! {\cal D} q\,
e^{
i\int\!\!dx\!\int\!dy\, \bar q(x) \bl[ (i \not\! \partial_x - m_q) \delta(x-y)
\,+\, \delta(x-y)\, g_0 \phi_0(y) \Gamma \br ]\,q(y)
}
\nn[2ex]
&=& {\rm det}|| (i \not\! \partial_x - m_q) \delta(x-y)
\,+\, \delta(x-y)\, g_0 \phi_0(y) \Gamma ||
\nn[2ex]
&\Longrightarrow&
 {\rm det}|| \delta(x-y)
\,+\, (i \not\! \partial_x - m_q)^{-1} \delta(x-y)\,
g_0 \phi_0(y) \Gamma ||
\nn[2ex]
&=&
{\rm det}|| \delta(x-y)
\,-\, i S_q(x-y) g_0 \phi_0(y) \Gamma ||
\nn[2ex]
&=&
\exp\bl\{
-\sum_{n=1}^\infty \frac{i^n}{n} g_0^n \int\!\! dx_1\ldots\!\int\!\! dx_n
\phi_0(x_1)\ldots\phi_0(x_n)
\Tr[\Gamma S_q(x_1-x_2)\ldots \Gamma S_q(x_n-x_1)]
\br\}.
\ena
one finds
\bea
Z_Y &=& \int\!\! {\cal D}\phi_0
\exp\bl\{
\frac{i}{2}\int\!\! dx \phi_0(x) (\Box - m^2_0) \phi_0(x)
\nonumber\\[2ex]
&-&
\sum_{n=1}^\infty \frac{i^n}{n} g_0^n \int\!\! dx_1\ldots\!\int\!\! dx_n
\phi_0(x_1)\ldots\phi_0(x_n)
\Tr[\Gamma S_q(x_1-x_2)\ldots \Gamma S_q(x_n-x_1)]
\br\}.
\label{eq:Yukawa_func2} 
\ena

Here we introduce the quark Green function in the usual form as
\be
 S_q(x-y) = \int\frac{d^4 k}{(2\pi)^4 i}\frac{e^{-ik(x-y)}}{m_q-\not\! k}\,.
\label{eq:prop}
\en

Then we collect the terms bi-linear in the boson fields.
One gets
\bea
L^{(2)}_Y &=& \frac12\!\int\!\! dx\, \phi_0(x) (\Box - m^2_0) \phi_0(x)
\nn
&-& \frac{i}{2}\,g_0^2\! \int\!\! dx_1\!\!\int\!\! dx_2
\phi_0(x_1)\phi_0(x_2)
\Tr[\Gamma S_q(x_1-x_2)\Gamma S_q(x_2-x_1)]
\nn[2ex]
&=&   \frac12\!\int\!\! dx\, \phi_0(x) (\Box - m^2_0) \phi_0(x)
        +\frac12\, g_0^2\! \int\!\! dx_1\!\!\int\!\! dx_2\,
\phi_0(x_1)\Pi_{S=0}(x_1-x_2)\phi_0(x_2)
\label{eq:L2}
\ena
where we define the mass function of the boson with spin $S=0$ as
\be
\Pi_{S=0}(x_1-x_2) = i\,\la T\bl\{
\bl[ \bar q \Gamma q\br]_x \bl[ \bar q \Gamma q \br]_y
                     \br\} \ra_0
= -i\, \Tr[\Gamma S_q(x_1-x_2)\Gamma S_q(x_2-x_1)]\,.
\label{eq:masfunS=0}
\en
Expanding the Fourier transform of this mass function
at the physical value of the boson mass up to the second order
\be
\tilde\Pi_{S=0}(p^2) = \int\!\! dx \,e^{-ipx}\Pi_{S=0}(x)
= \tilde\Pi_{S=0}(m^2) + (p^2 - m^2)  \tilde\Pi'_{S=0}(m^2)
 + \tilde\Pi_{S=0}^{\rm ren}(p^2)
\label{eq:FourierS=0}
\en
one finds
\bea
L^{(2)}_Y &=& \frac12\!\int\!\! dx\,
\phi_0(x) (\Box - m^2_0 + g_0^2  \tilde\Pi_{S=0}(m^2)
+(\Box - m^2) \tilde\Pi'_{S=0}(m^2)) \phi_0(x)
\nn
&+& \frac12\, g_0^2\! \int\!\! dx_1\!\!\int\!\! dx_2\,
\phi_0(x_1)\Pi^{\rm ren}_{S=0}(x_1-x_2)\phi_0(x_2)\,.
\label{eq:L2ren}
\ena
The renormalization of boson mass, wave function and
Yukawa coupling proceed in the  standard manner
\bea
m^2 &=& m^2_0 - g_0^2\,\tilde\Pi_{S=0}(m^2), \qquad \phi_r=Z^{-1/2}\phi_0,
\nn[2ex]
g_r &=& Z^{1/2} g_0, \qquad Z=\frac{1}{1+g_0^2\,\tilde\Pi'_{S=0}(m^2)}\,.
\label{eq:renorm}
\ena
Note that the wave function renormalization constant may be expressed
via the renormalized coupling constant
\be
Z= 1 - g_r^2\,\tilde\Pi'_{S=0}(m^2)\,.
\label{eq:Zr}
\en
Finally, the renormalized  generating functional for the Yukawa theory
is written as
\bea
Z^{\rm ren}_Y &=& \int\!\! {\cal D}\phi_r
\exp\bl\{
\frac{i}{2}\int\!\! dx \phi_r(x) (\Box - m^2) \phi_r(x)
\nonumber\\
&+& \frac{i}{2}\, g_r^2\! \int\!\! dx_1\!\!\int\!\! dx_2\,
\phi_r(x_1)\Pi^{\rm ren}_{S=0}(x_1-x_2)\phi_r(x_2)
\nn
&-&
\sum_{n=3}^\infty \frac{i^n}{n} g_r^n \int\!\! dx_1\ldots\!\int\!\! dx_n
\phi_r(x_1)\ldots\phi_r(x_n)
\Tr[\Gamma S_q(x_1-x_2)\ldots \Gamma S_q(x_n-x_1)]
\br\}\,. \label{eq:Yukawa_ren}
\ena
We drop the linear boson term because it is absent for
pseudoscalar mesons and it can be removed in the scalar case by 
a shift of the field.

Now we consider the  generating functional for the Fermi theory
\be
Z_F = \int\!\! {\cal D}\bar q\! \int\!\! {\cal D} q\,
e^{i\int\! dx {\cal L}_F(x)}
\label{eq:Fermi_func1}
\en
where the Lagrangian ${\cal L}_F(x)$ is given by Eq.~(\ref{eq:Fermi}).
By using the Gaussian functional representation for the exponential
of the four-fermion interaction
\be
e^{i\frac{G}{2} \la (\bar q\Gamma q)^2 \ra}
= N_+^{-1} \int\!\! {\cal D}\phi
\exp\{-\frac{i}{2}\frac{1}{G}\la\phi^2\ra
+i\la \phi\cdot (\bar q\Gamma q )\ra\} \quad\text{where}\quad
\la(...)\ra = \int\! dx\, (...)
\label{eq:four-fermion}
\en
one finds
\bea
Z_F =  N_+^{-1} \int\!\! {\cal D}\phi\!
               \int\!\! {\cal D}\bar q\! \int\!\! {\cal D} q\,
\exp\{
-\frac{i}{2}\frac{1}{G}\la\phi^2\ra
\,+\,
i\,\la \bar q \bl[ i \not\! \partial - m_q \,+\,
\phi \Gamma \br] q \ra
\}.
\label{eq:Fermi_func2}
\ena
Then we integrate out the quark fields
\bea
Z_F &=& \int\!\! {\cal D}\phi
\exp\bl\{
- \frac{i}{2} \frac{1}{G}\int\!dx\,\phi^2(x)
\nonumber\\[2ex]
&-&
\sum_{n=1}^\infty \frac{i^n}{n}  \int\!\! dx_1\ldots\!\int\!\! dx_n
\phi(x_1)\ldots\phi(x_n)
\Tr[\Gamma S_q(x_1-x_2)\ldots \Gamma S_q(x_n-x_1)]
\br\}\label{eq:Fermi_func3}
\ena
where we drop all irrelevant normalization constants.
Next we collect the terms bi-linear in the boson fields
and introduce the renormalized mass function like
in the Yukawa case. One gets
\bea
L^{(2)}_F &=& \frac12\!\int\!\! dx\,
\phi(x) \Big(-\frac{1}{G} + \tilde\Pi_{S=0}(m^2)
+ (\Box - m^2) \tilde\Pi'_{S=0}(m^2)\Big) \phi(x)
\nn
&+& \frac12\,\int\!\! dx_1\!\!\int\!\! dx_2\,
\phi(x_1)\Pi^{\rm ren}_{S=0}(x_1-x_2)\phi(x_2)\,.
\label{eq:L2Fren}
\ena
If we require the condition
\be
G\,\tilde\Pi_{S=0}(m^2)=1
\label{eq:mass}
\en
and rescale the boson field as $\phi\to\phi/\sqrt{\tilde\Pi'_{S=0}(m^2)}$
one obtains the free Lagrangian of the  boson field with
the mass $m$ and the correct residue of the Green function.
The fully renormalized generating functional of the Fermi-theory is written as

\bea
Z^{\rm ren}_F &=& \int\!\! {\cal D}\phi
\exp\bl\{
\frac{i}{2}\int\!\! dx \phi(x) (\Box - m^2) \phi(x)\nonumber\\
&+& \frac{i}{2}\, \frac{1}{\tilde\Pi'_{S=0}(m^2)}\!
\int\!\! dx_1\!\!\int\!\! dx_2\,
\phi(x_1)\Pi^{\rm ren}_{S=0}(x_1-x_2)\phi(x_2)
\nn
&-&
\sum_{n=3}^\infty \frac{i^n}{n}
\left[ \frac{1}{\sqrt{\tilde\Pi'_{S=0}(m^2)}} \right]^n
\int\!\! dx_1\ldots\!\int\!\! dx_n\phi(x_1)\ldots\phi(x_n)
\nn
&\times&
\Tr[\Gamma S_q(x_1-x_2)\ldots \Gamma S_q(x_n-x_1)]
\br\}\,.
\label{eq:Fermi_ren} 
\ena
Comparing both renormalized generating functionals 
of Eqs.~(\ref{eq:Yukawa_ren}) 
and~(\ref{eq:Fermi_ren}) we conclude that the condition for their equality is
\be
g_r = \frac{1}{\sqrt{\tilde\Pi'_{S=0}(m^2)}}
\en
or, according to  Eq.~(\ref{eq:Zr}),
\be
Z= 1 - g_r^2\,\tilde\Pi'_{S=0}(m^2)=0\,.
\label{eq:Z=0}
\en
Thus the vanishing of the wave function renormalization constant
in the Yukawa theory may be interpreted as the condition
that the bare, unrenormalized field $\phi_0=Z^{1/2}\phi_r$
vanishes for a composite boson. As follows from Eq.~(\ref{eq:renorm})
the bare boson mass $m_0$ and the bare Yukawa coupling $g_0$
go to infinity when the renormalization constant $Z$ goes to zero.
But the limit proceeds in such a way that
\be
\frac{g_0^2}{\delta m^2} \underbrace{=}_{g_0^2,m_0^2\to\infty}  G<\infty
\qquad \delta m^2\equiv  m_0^2-m^2.
\en
This limiting process is called the Jouvet condition \cite{Jouvet:1956ii}.

\section{Mass function in the covariant quark model}

The interaction of the ground-state pseudoscalar and vector mesons
with their constituent quarks is described in the covariant quark model
by a Lagrangian which reads
\bea
{\cal L}_{\rm int} &=& g_H H(x)\,J_H(x)\,; \qquad
J_H(x) = \int\!\! dx_1 \!\!\int\!\!dx_2 F_H (x;x_1,x_2)
\bar q_2(x_2) \Gamma_H q_1(x_1)\,.
\label{eq:int}
\ena
Here, $\Gamma_P=i\gamma^5$ and $\Gamma_V^\mu=\gamma^\mu$
are chosen for the  pseudoscalar and vector mesons, respectively.
The vector meson field $\phi^\mu$ has the Lorentz index $\mu$ and
satisfies the transversality condition
\be
\partial_\mu\phi^\mu = 0\,.
\label{eq:trans}
\en
For the vertex function $F_H$ we use the translationally invariant form
\be
F_H(x,x_1,x_2)=\delta(x - w_1 x_1 - w_2 x_2) \Phi_H((x_1-x_2)^2)
\label{eq:vertex}
\en
where $w_i = m_{q_i}/( m_{q_1}+ m_{q_2} )$ so that $w_1+w_2=1$.
The Fourier transform of the vertex function is chosen in a
Gaussian form
\be
\tilde\Phi_H(-p^2) = \int\! dx\, e^{ipx} \Phi_H(x^2)
= e^{p^2/\Lambda^2_H} 
\label{eq:Gauss}
\en
for both the pseudoscalar and vector mesons. The size parameter
$\Lambda_H$ is an adjustable quantity. Since the calculation of
the Feynman diagrams proceeds in the Euclidean region where
$p^2=-p^2_E$, the vertex function decreases very rapidly for
$p^2_E\to\infty$ and thereby provides ultraviolet convergence
in the evaluation of any diagram.

The mass functions for the pseudoscalar (spin $S=0$)
and vector mesons (spin $S=1$) are defined as
\bea
\Pi_{PP}(x-y) &=& +\,i\,\la T\bl\{J_P(x)J_P(y) \br\} \ra_0 ,
\label{eq:S=0}\\[2ex]
\Pi^{\mu\nu}_{VV}(x-y) &=& -\,i\,\la T\bl\{J^\mu_V(x)J^\nu_V(y) \br\} \ra_0 
\,. \label{eq:S=1}
\ena
Using the Fourier transforms of the vertex functions of Eq.~(\ref{eq:Gauss})
and of the quark propagators with Eq.(\ref{eq:prop}) one can easily find
the Fourier transforms of the mass functions
\bea
\tilde\Pi_{PP}(p^2) &=& N_c\int\frac{d^4k}{(2\pi)^4i} \tilde\Phi^2_P(-k^2)
\Tr\Big(\gamma^5 S_1(k+w_1 p)\gamma^5 S_2(k-w_2 p)\Big),
\label{eq:massP-1}\\[2ex]
\tilde\Pi^{\mu\nu}_{VV}(p) &=& N_c\int\frac{d^4k}{(2\pi)^4i} 
\tilde\Phi^2_V(-k^2)
\Tr\Big(\gamma^\mu S_1(k+w_1 p)\gamma^\nu S_2(k-w_2 p)\Big)
\nonumber\\ 
&=& g^{\mu\nu} \tilde\Pi_{VVg}(p^2) + p^\mu p^\nu  \tilde\Pi_{VVpp}(p^2)
\label{eq:massV-1}
\ena
where $N_c=3$ is a number of color degrees of freedom. 
Due to the transversality of the
vector field the second term in Eq.~(\ref{eq:massV-1}) is irrelevant
in our considerations. The first remaining term 
in Eq.~(\ref{eq:massV-1}) can be expressed as
\be
 \tilde\Pi_{VVg}(p^2) = \frac13 \bl(g_{\mu\nu}-\frac{p_\mu p_\nu}{p^2} \br)
                       \tilde\Pi^{\mu\nu}_{VV}(p)\,.
\label{eq:massV-2}
\en

By using the calculational technique from Ref.~\cite{Branz:2009cd}
one finds
\bea
\tilde\Pi_H(p^2)&=& \frac{3}{4\pi^2}\int\limits_0^{1/\lambda^2} \!\!
\frac{dt\,t}{a_H^2} \int\limits_0^1\!\!d\alpha\,
e^{-t\,z_0 + z_H}\,
\Big\{
  \frac{n_H}{a_H} + m_{q_1}m_{q_2}
+ \bl(w_1 - \frac{b}{a_H}\br) \bl(w_2 + \frac{b}{a_H}\br)p^2
\Big\}\,, 
\label{eq:mass_fin}
\ena
where 
\bea 
z_0 &=&  \alpha m^2_{q_1} +(1-\alpha)m^2_{q_2} - \alpha(1-\alpha) p^2\,,
\qquad z_H  = \frac{2s_Ht}{2s_H+t} (\alpha-w_2)^2 p^2 \,,
\nonumber\\
a_H &=& 2s_H+t\, , \qquad b = (\alpha-w_2)t\,. 
\ena
Here $n_P=2$ and $n_V=1$. We use the result of the fit \cite{Ivanov:2011aa}
for the value of infrared cutoff with $\lambda=181$~MeV. The parameter $s_H$
is related to  the size parameter  $\Lambda_H$  as $s_H=1/\Lambda^2_H$.
Note that in the case  $\lambda\to 0$ the branching point appears
at $p^2=(m_{q_1} + m_{q_2})^2$. At this point the integral over $t$ becomes
divergent as $t\to\infty$ because of $z_0=0$
at $\alpha=m_{q_2}/ (m_{q_1} + m_{q_2})$.  By introducing an infrared cutoff
on the upper limit of the scale integration one can avoid the appearance
of the threshold singularity.

The compositeness condition
\be
Z_H = 1- g^2_H \,\tilde\Pi'_H(m^2_H) = 0\,,
\label{eq:Z=0_model}
\en
where $g_H$ is the already renormalized Yukawa coupling constant,
now has a clear mathematical meaning because the mass function
$\Pi_H$ in Eq.~(\ref{eq:mass_fin}) is well defined.

As discussed in the previous section, the Yukawa theory defined
by the interaction Lagrangian of Eq.~(\ref{eq:int}) is equivalent
to the Fermi theory defined by  the interaction Lagrangian
\be
{\cal L}^F_{\rm int} = \frac{G}{2} J_H^2(x)
\label{eq:int_Fermi}
\en
if the wave function renormalization constant $Z_H$ is equal to zero 
and the Fermi coupling $G$ satisfies the equation 
\be
G\,\tilde\Pi_H(m^2_H) = 1\,.
\label{eq:Fermi_coupl}
\en
Now we are able to investigate the dependence of the Fermi coupling $G$
on the hadron masses.

\section{Numerical results}

A first fit of the model parameters has originally been performed in
Ref.~\cite{Branz:2009cd}, where the above described method for implementing
infrared quark confinement was used for the first time.
The leptonic decay constants which are known either from experiment
or from lattice simulations  have been chosen as input quantities
to adjust the model parameters. A given meson $H$ in the interaction
Lagrangian Eq.~(\ref{eq:int}) is characterized by the coupling constant $g_H$,
the size parameter $\Lambda_H$ and two of the four constituent quark masses,
$m_q$ ($m_u=m_d$, $m_s$, $m_c$, $m_b$). Moreover, there is the infrared
confinement parameter $\lambda$ which is universal for all hadrons.
Note that the physical values for the hadron masses have been used
in the fit. In the beginning we have $2n_H+5$ adjustable parameters for $n_H$
number of mesons. The compositeness condition (\ref{eq:Z=0_model}) provides
$n_H$ constraints and allows one to express all coupling constants $g_H$
through other model parameters. 
The remaining  $n_H+5$ parameters are determined by
a fit to experimental data.  As input data the values of the leptonic decay
constants and some electromagnetic decay widths are chosen.
Later on, several updated fits were indicated in Ref.~\cite{Ivanov:2011aa}.
In this paper we will use one of them which is slightly different from
the published version. The reason is that in the published 
version~\cite{Ivanov:2011aa} the value of the charm quark mass was found
to be $m_c=2.16$ GeV which is a somewhat higher than the value
needed to describe some observables of in the charm sector.
The results of the (overconstrained) least--squares fit
used in the present study  can be found in
Tables~\ref{tab:leptonic} and Table~\ref{tab:em-widths}. 
The agreement between the fit and input values is
quite satisfactory. We do not include decay results for
the $\eta(\eta')$-mesons because the primary goal of our present study
is to understand the origin of the meson masses in the framework of
the covariant quark model.
The  $\eta(\eta')$-mesons have the additional features like the mixing angle
and an possibly important gluon admixture to the conventional 
$q\bar q$-structure of the $\eta'$.
Some aspects of the nonleptonic $B_s$-meson decays with $\eta(\eta')$
in the final states were recently discussed in Ref.~\cite{Dubnicka:2013vm}.
The results of the fit for the values of the quark masses $m_{q_{i}}$, the
infrared cutoff parameter $\lambda$ and the size parameters $\Lambda_{H_{i}}$
are given in (\ref{eq: fitmas}) and in Table~\ref{tab:Lambda-H}, respectively.
The constituent quark masses and the values for the size parameters
fall into the expected range. The size parameters show the expected general
pattern: the geometrical size of a meson, which is inversely
proportional to $\Lambda_{H_{i}}$, decreases when the mass increases. 

The present numerical least-squares fit and the values for the model
parameters supersede the results of a similar analysis
given in \cite{Branz:2009cd}, where a different set of
electromagnetic decays has been used.
In the present fit we have also updated some of the theoretical/experimental
input values.

\begin{table}[ht]
\caption{Input values for the leptonic decay constants $f_H$ (in MeV) and
our least-squares fit values.}
\label{tab:leptonic}
\begin{center}
\def\arraystretch{1.2}
\begin{tabular}{ccll}
\hline\hline
    & Fit Values  & Data &  Ref.  \\
\hline
$f_\pi$  & 128.4 & $130.4 \pm 0.2 $   & \cite{PDG,RosnerStone}\\
$f_K$   & 156.0 & $156.1 \pm 0.8 $  & \cite{PDG,RosnerStone}\\
$f_{D}$  & 206.7 & $206.7 \pm 8.9 $ & \cite{PDG,RosnerStone}\\
$f_{D_s}$ & 257.5 & $257.5 \pm 6.1 $ & \cite{PDG,RosnerStone}\\
$f_{B}$ & 189.7 & $192.8 \pm 9.9  $ & \cite{LatticePRD81}\\
$f_{B_s}$ & 235.3 & $238.8 \pm 9.5 $ & \cite{LatticePRD81}\\
$f_{\eta_c}$ & 386.6 & $438 \pm 8 $ & \cite{LatticeTWQCD}\\
$f_{B_c}$ & 445.6 & $489 \pm 5 $ & \cite{LatticeTWQCD}\\
$f_{\eta_b}$ & 609.1 & $801 \pm 9 $ & \cite{LatticeTWQCD}\\
\hline
$f_{\rho}$ & 221.2 & $221 \pm 1 $ & \cite{PDG}\\
$f_\omega$   & 204.2  & $198 \pm 2 $ & \cite{PDG} \\
$f_\phi$     & 228.2  & $227\pm  2 $ & \cite{PDG} \\
$f_{\Jpsi}$   & 415.0  & $415\pm 7  $ & \cite{PDG} \\
$f_{K^\ast}$   & 215.0  & $217\pm 7  $ & \cite{PDG} \\
$f_{D^\ast}$   & 223.0  & $245\pm 20  $ & \cite{Lubicz} \\
$f_{D^\ast_s}$  & 272.0  & $272\pm26  $ & \cite{Lubicz} \\
$f_{B^\ast}$   & 196.0  & $196\pm 44 $ & \cite{Lubicz} \\
$f_{B_s^\ast}$  & 229.0  & $229\pm 46  $ & \cite{Lubicz} \\
$f_{\Upsilon}$  & 661.3  & $715\pm 5  $ & \cite{PDG} \\
\hline\hline
\end{tabular}
\end{center}
\end{table}

\begin{table}[ht]
\begin{center}
\def\arraystretch{1.2}
\caption{Input values for some basic electromagnetic decay widths and our
least-squares  fit values (in keV).}
\label{tab:em-widths}
\vspace*{0.2cm}
\begin{tabular}{lll}
\hline\hline
Process & Fit Values & Data~\cite{PDG}  \\
\hline
$\pi^0\to\gamma\gamma$          & \,\,  $ 5.07 \times 10^{-3}$ \,\,&
                                \,\,$(7.7 \pm 0.4) \times 10^{-3}$\,\,\\
$\eta_c\to\gamma\gamma$         & 3.47 & 5.0 $\pm$ 0.4 \\
$\rho^{\pm}\to\pi^{\pm}\gamma$    & 76.3     &  67 $\pm$ 7  \\
$\omega\to\pi^0\gamma$          & 687      &  703 $\pm$ 25    \\
$K^{\ast \pm}\to K^\pm\gamma$      & 57.7     &  50 $\pm$ 5          \\
$K^{\ast 0}\to K^0\gamma$         & 129      &  116 $\pm$ 10      \\
$D^{\ast \pm}\to D^\pm\gamma$      & 0.59     &  1.5 $\pm$ 0.5 \\
$\Jpsi \to \eta_c \gamma $      & 1.90     &  1.58 $\pm$ 0.37  \\
\hline
\end{tabular}
\end{center}
\end{table}

\be
\def\arraystretch{1.5}
\begin{array}{cccccc}
     m_{u/d}        &      m_s        &      m_c       &     m_b & \lambda  &
\\\hline
 \ \ 0.235\ \   &  \ \ 0.442\ \   &  \ \ 1.61\ \   &  \ \ 5.07\ \   &
\ \ 0.181\ \   & \ {\rm GeV}
\end{array}
\label{eq: fitmas}
\en

\begin{table}[ht]
\caption{The fitted values of the size parameters  $\Lambda_H$ in GeV.}
\label{tab:Lambda-H}
\begin{center}
\def\arraystretch{1.5}
\begin{tabular}{cccccccccc}
\hline
 $\pi$ & $K$  & $D$  & $D_s$ & $B$  & $B_s$ & $B_c$ & $\eta_c$ & $\eta_b$ &\\
  0.87 & 1.02 & 1.71 & 1.81  & 1.90 & 1.94  & 2.50 & 2.06 & 2.95 &\\
\hline\hline
 $\rho$ & $\omega$ & $\phi$ & $\Jpsi$ & $K^\ast$ & $D^\ast$ & $D_s^\ast$ &
  $B^\ast$ & $B_s^\ast$ & $\Upsilon$ \\
 \ \ 0.61\ \  & \ \  0.50 \ \  & \ \  0.91\ \  & \ \  1.93\ \  & \ \  0.75\ \
 & \ \  1.51\ \  & \ \  1.71 \ \  & \ \  1.76\ \  & \ \  1.71 \ \ &
\ \ 2.96 \ \ \\
\hline
\end{tabular}
\end{center}
\end{table}

\begin{figure}[htp]
\begin{center}
\includegraphics[scale=.5]{GF.eps}
\end{center}
\caption{\label{fig:G_Fermi}
The dependence of the Fermi coupling on the fitted meson masses.}

\begin{center}
\vspace*{2cm}
\includegraphics[scale=1]{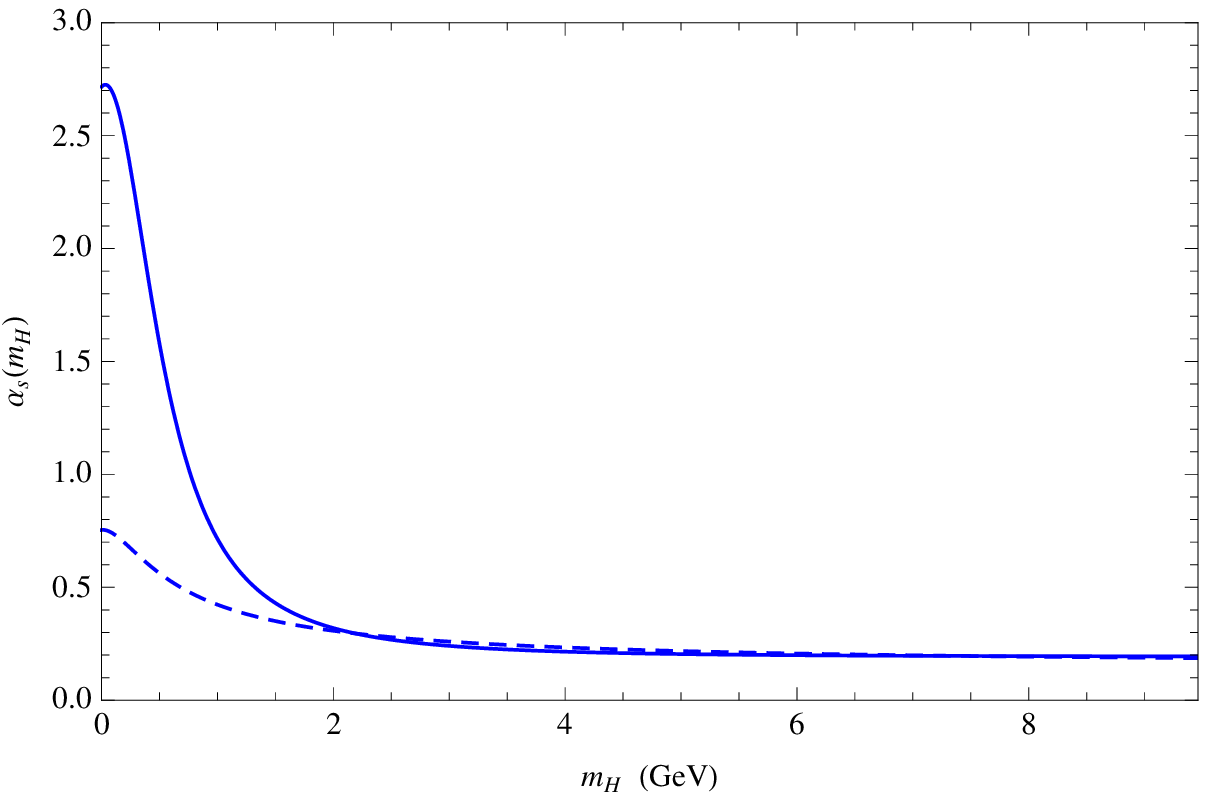}
\end{center}
\caption{\label{fig:alpha_S}
The comparison of the effective coupling $\alpha_s(m_H)$ obtained in
\cite{Ganbold:2010bu,Ganbold:2012zz} (dashed line) with our calculated
dimensionless coupling $\lambda^2 G(m_H)$  (solid line) scaled to the
curve $\alpha_s(m_H)$  in the region of large masses.}
\end{figure}

Our prime goal is to study the behavior of the Fermi coupling $G$
in Eq.~(\ref{eq:Fermi_coupl}) as a function of the hadron masses
by keeping other parameters (infrared cutoff parameter $\lambda$,
size parameters $\Lambda_H$ and constituent quark masses $m_q$)
fixed. The original dependence of $G$ on the hadron mass is
obtained by directly taking the physical values, resulting in 
a sawtooth-like behavior.
We therefore suggest to change the values of the input hadron masses in such a
way to get a relatively  smooth dependence of $G$ on the masses.
{\it A smoothness criterion} might be considered as a possibility,
when values for the meson masses are computed through
Eq.~(\ref{eq:Fermi_coupl}) as a function of the other model parameters. 
The obtained smooth dependence of the dimensionless quantity $G\lambda^2$
on these masses is shown in Fig.~\ref{fig:G_Fermi} where the calculated
values are connected by straight lines.
The estimated values for the meson masses found in this manner
are shown in Table~\ref{tab:masses}. One can see that
they are in quite good agreement with the experimental data. 
For completeness in Table~\ref{tab:Glambda2} we also present our results 
for the effective couplings $G\lambda^2$ in the case of exact fit 
(when the values of meson masses are taken from data) and in 
the case of the smooth fit. 

\begin{table}[ht]
\caption{The fitted  values for the meson masses in MeV}
\label{tab:masses}
\begin{center}
\def\arraystretch{1.2}
\begin{tabular}{lll}
\hline\hline
    & Model  & \quad Data \cite{PDG}  \\
\hline
$m_\pi$      & 141.0  & \quad 139.57018 $\pm$ 0.0003 \\
$m_K$       & 493.0  & \quad 493.677 $\pm$ 0.016  \\
$m_{\rho}$   & 778.0  & \quad 775.26 $\pm$ 0.25 \\
$m_\omega$   & 806.0  & \quad 782.65 $\pm$ 0.12 \\
$m_{K^\ast}$  & 893.0  & \quad 891.66  $\pm$  0.26\\
$m_\phi$    & 1011.0  & \quad 1019.45 $\pm$ 0.02  \\
$m_{D}$     & 1915.0  & \quad 1869.62 $\pm$ 0.15 \\
$m_{D_s}$    & 1998.0  & \quad 1968.50  $\pm$ 0.32 \\
$m_{D^\ast}$  & 2001.0  & \quad 2010.29  $\pm$ 0.13 \\
$m_{D^\ast_s}$ & 2099.0  & \quad 2112.3 $\pm$ 0.5 \\
$m_{\eta_c}$  & 2922.0  & \quad 2983.7 $\pm$ 0.7  \\
$m_{\Jpsi}$   & 3067.0  &  \quad 3096.916 $\pm$ 0.011 \\
$m_{B}$      & 5425.0  & \quad  5279.26  $\pm$ 0.17 \\
$m_{B^\ast}$  & 5450.0   & \quad 5325.2 $\pm$  0.4 \\
$m_{B_s}$    & 5524.0  & \quad 5366.77  $\pm$ 0.24\\
$m_{B_s^\ast}$ & 5566.0  & \quad 5415.8  $\pm$ 1.5 \\
$m_{B_c}$    & 6041.0  & \quad 6274.5 $\pm$ 1.8 \\
$m_{\eta_b}$    & 8806.0  & \quad 9398.0 $\pm$ 3.2 \\
$m_{\Upsilon}$  & 8880.0  & \quad 9460.30 $\pm$ 0.26 \\
\hline\hline
\end{tabular}
\end{center}
\end{table}

\begin{table}[ht]
\caption{Values for effective couplings $G \lambda^2$ in 
cases of exact and smooth fit} 
\label{tab:Glambda2}
\begin{center}
\def\arraystretch{1.2}
\begin{tabular}{lll}
\hline\hline
         \ \ & Exact fit  \ \ & Smooth fit \\ 
\hline
$\pi$    & 1.508      &  1.507  \\      
$K$      & 0.919      &  0.920  \\      
$\rho$   & 0.571      &  0.560  \\      
$\omega$ & 0.673      &  0.553  \\      
$K^\ast$ & 0.476      &  0.472 \\
$\phi$   & 0.377      &  0.400  \\
$D$      & 0.224      &  0.195  \\
$D_s$    & 0.197      &  0.184  \\
$D^\ast$ & 0.168      &  0.180  \\
$D^\ast_s$& 0.158     &  0.170  \\
$\eta_c$  & 0.128     &  0.141  \\
$\Jpsi$   & 0.129     &  0.139  \\ 
$B$       & 0.215     &  0.125 \\
$B^\ast$  & 0.237     &  0.124 \\
$B_s$     & 0.192     &  0.122 \\
$B_s^\ast$& 0.232     &  0.121 \\ 
$B_c$     & 0.0905    &  0.118 \\
$\eta_b$  & 0.0612    &  0.0986\\  
$\Upsilon$& 0.0600    &  0.0984\\  
\hline\hline
\end{tabular}
\end{center}
\end{table}

It might be interesting to compare the behavior of $G$ with the effective
QCD coupling constant $\alpha_s$ obtained in
the relativistic models with specific forms of analytically confined
quark and gluon propagators
\cite{Ganbold:2009ak,Ganbold:2010bu,Ganbold:2012zz}.
In these models the nonlocal four-quark interaction is induced
by one-gluon exchange between biquark currents. Since the quark currents
are connected via the confined gluon propagator having the dimension
of an inverse mass squared in momentum space, the resulting coupling
$\alpha_s$ is dimensionless. 
In Fig.~\ref{fig:alpha_S}  we compare the mass dependence of
the  rescaled dimensionless Fermi coupling 
$\alpha_s^{\rm model}\equiv 1.74 \, G\lambda^2$ [solid line] 
estimated for the model parameters given by Eq.~(\ref{eq: fitmas}) with
the effective QCD coupling $\alpha_s$ [dashed line] obtained 
in~\cite{Ganbold:2010bu,Ganbold:2012zz}. 
The idea of such a comparison is to check for identical functional 
behavior, even when a rescaling is involved. 
Here, for $m_H > 2$ GeV both curves agree rather well, which is nontrivial 
information. After rescaling we are able to compare the behavior of the 
two curves in the region of small masses. 
They are different due to different
dynamics (confinement, quark propagators, vertex functions, etc.)
implemented in these approaches. 
Note, the particular choice of the model parameters used in
Ref.~\cite{Ganbold:2010bu} are
$m_{u/d}=0.193, m_s=0.293, m_c=1.848, m_b=4.693$~GeV  for the constituent
quark masses and $\Lambda=0.345$ GeV for the confinement scale. 
Despite the different model origins and input parameter values, 
the behaviors of two curves  are very similar to each other
in the intermediate and heavy mass regions above $\sim 2$ GeV.
Their values at the origin are mostly determined by the confinement
mechanisms realized in different ways in these models.
This could explain why they have  different
behaviors  in the low-energy region below 2 GeV. 

\section{Summary}

We have represented a brief sketch of an approach to the bound state problem
in quantum field theory which is based on the compositeness condition $Z_H=0$.
By using the functional integral we have demonstrated explicitly
that the four-fermion theory with the Fermi coupling $G$ is equivalent
to the Yukawa-type theory if, first, the wave function renormalization
constant in the Yukawa theory is equal to zero and, second,
the Fermi coupling $G$  is inversely proportional
to the meson mass function calculated for the physical meson mass.

We have given details for the calculation of the mass function for
pseudoscalar and vector mesons in the framework of the covariant quark model.
We updated the fit of the model parameters and calculated the 
Fermi coupling $G$ as a function of physical masses in 
a quite large region from the $\pi$ up to $B_c$ mesons.

We have suggested {\it a smoothness criterion} for the curve
just varying the meson masses in such a way to obtain
the smooth behavior for the Fermi coupling $G$. The mass spectrum obtained
in this manner is found to be in good agreement with the experimental data.
We have compared the behavior of $G$ with the strong QCD coupling $\alpha_s$
calculated in the QCD-inspired approach.

\begin{acknowledgments}

This work was supported by the DFG under Contract No. LY 114/2-1
and by Tomsk State University Competitiveness Improvement Program.
G.G.  gratefully acknowledges support from the Alexander von Humboldt
Foundation and would like to thank Institut f\"ur Theoretische Physik,
Universit\"at T\"ubingen for warm hospitality.  M.A.I.\ acknowledges
the support from Mainz Institute for Theoretical Physics (MITP) and
the Heisenberg-Landau Grant.

\end{acknowledgments}

\end{document}